\documentclass[aps,pra,twocolumn,showpacs,groupedaddress]{revtex4}
\usepackage{graphicx}
\usepackage{amsmath}
\usepackage{longtable}
\usepackage{afterpage}
\usepackage{rotating}
\begin{document}

\title{Black-body radiation shift of the Ga$^{+}$ clock transition}
\author{Yongjun Cheng$^{1,2}$ and J. Mitroy$^{2}$ } 
\affiliation {$^1$The Academy of Fundamental and Interdisciplinary Science, Harbin Institute of Technology, Harbin 150080, People$'$s Republic of China}
\affiliation {$^2$School of Engineering, Charles Darwin University, Darwin NT 0909, Australia} 

\date{\today}

\begin{abstract}
The blackbody radiation shift of the Ga$^+$ $4s^2 \ ^1S^e_0 \to 4s4p \ ^3P^o_0$ 
clock transition is computed to be $-$$0.0140 \pm 0.0062$ Hz at 300 K.  The
small shift is consistent with the blackbody radiation shifts of the clock 
transitions of other group III ions which are of a similar size.  The 
polarizabilities of the Ga$^+$ $4s^2 \ ^1S^e_0$, $4s4p \ ^3P^o_0$, and 
$4s4p \ ^1P^o_1$ states were computed using the configuration interaction 
method with an underlying semi-empirical core potential.  Quadrupole and 
non-adiabatic dipole polarizabilities were also computed.  
A byproduct of the analysis involved calculations of the low 
lying spectrum and oscillator strengths, including polarizabilities, of 
the Ga$^{2+}$ ion. 

\end{abstract}

\pacs{32.10.Dk, 31.15.ap, 31.15.V-, 32.70.Cs}

\maketitle

\section{Introduction} 

The current standard of time is based on the cesium 
fountain frequency standard \cite{gill05a,margolis09a}. However, 
recent developments in cold atom physics and improvements in optical 
frequency measurements make it increasingly likely that an atomic
clock based on an optical transition will supplant the current 
cesium standard and consequently lead to a new definition of the 
second \cite{gill11a}.  At present, the smallest frequency uncertainty 
has been achieved by an optical clock based on quantum logic technology 
and using the highly forbidden $3s^2 \ ^1S^e_0$ $\to$ $3s3p \ ^3P^o_0$ 
transition.  The fractional frequency uncertainty of this clock is 
only $8.6 \times 10^{-18}$ \cite{rosenband08a,chou10a}.  This clock would 
only drift by a period of 1 second over a period of $3.7 \times 10^9$ years.

Optical frequency standards capable of achieving such extreme precisions 
are however sensitive to very small environmental influences.   One of 
the most important of these influences is blackbody radiation (BBR) 
emitted by the apparatus containing the atomic or ionic clock.  The 
electromagnetic field associated with this blackbody radiation results 
in an AC Stark shift of the energies of the two states that define 
the clock transition.  The energies of the upper and lower states of the 
clock transition can shift by different amounts since the polarizabilities 
of the two levels will not necessarily be the same.   This leads to a 
temperature dependent shift in the frequency of the clock 
\cite{margolis03a,porsev06b,mitroy10a}.  It is expected that the BBR shift 
will become an increasingly important component of the error budgets for 
optical frequency 
standards as other potential sources of uncertainty are eliminated and their 
overall precision is improved \cite{hunag11b,middlemann12a,sherman12a}.   Consequently, 
clock transitions involving 
upper and lower states having polarizabilities that are close to each other are 
attractive 
since they will have small BBR shifts.  Indeed, the very small BBR shift 
\cite{mitroy09b} was a primary motivation for the development of the Al$^+$ 
frequency standard \cite{rosenband06a,rosenband07a,rosenband08a,chou10a}.  Just 
recently, investigations of the $ns^2 \ ^1S^e_0 \to nsnp \ ^3P^o_0$ transitions 
of other group III positive ions of the periodic table, 
namely, B$^+$, In$^+$ and Tl$^+$ have been completed.  The BBR shifts of 
all these ions were found to be small, with In$^+$ having the largest 300 K BBR 
shift of $-$0.017 Hz \cite{safronova11c,zuhrianda12a,cheng12a}.  
All of these BBR shifts were between 1 to 2 orders of
magnitude smaller than the BBR shifts of other atoms and ions advanced as 
atomic frequency standards \cite{mitroy10a}.  Therefore, it is somewhat surprising 
that there has not yet been a calculation of the BBR shift of the Ga$^+$ clock 
transition.    

Calculations of the polarizabilities of the Ga$^+$ $4s^2 \ ^1S^e_0$,  
$4s4p \ ^3P^o_0$ and $4s4p \ ^1P^o_1$ states have been made 
using a large configuration interaction (CI) calculation to account 
for valence correlation. Core-valence correlations were included 
by adding semi-empirical core-polarization potentials to the core 
potential based on a Hartree-Fock wave function for the Ga$^{3+}$ 
core. The BBR shift is calculated and found to be small and roughly 
the same size as the shifts for other group III ions.   

\section{Structure calculations} 

The CI calculations used to generate the physical and $L^2$ pseudo
states were similar in style to those used previously to determine
the dispersion parameters and polarizabilities of a number of two
electron systems \cite{mitroy03f,mitroy04b,mitroy08k,mitroy09b}.
The Hamiltonian for the two active electrons is written
\begin{eqnarray}
H  &=&  \sum_{i=1}^2 \left(  -\frac {1}{2} \nabla^2_i
 + V_{\rm dir}({\bf r}_i) + V_{\rm exc}({\bf r}_i) +  V_{\rm p1}({\bf r}_i) \right) \nonumber \\
 &+&  V_{\rm p2}({\bf r}_1,{\bf r}_2) + \frac{1}{r_{12}} \ .
\end{eqnarray}
The direct, $V_{\rm dir}$, and exchange, $V_{\rm exc}$, interactions of
the valence electrons with the Hartree-Fock (HF) core were calculated 
exactly. The $1s^22s^22p^63s^23p^63d^{10}$ core wave function was taken 
from a HF calculation of the Ga$^{2+}$ ground state using a Slater type 
orbital (STO) basis. The $\ell$-dependent polarization potential, 
$V_{\rm p1}$, was semi-empirical in nature with the functional form
\begin{equation}
  V_{\rm p1}({\bf r})  =  -\sum_{\ell m} \frac{\alpha_{\rm core} g_{\ell}^2(r)}{2 r^4}
           |\ell m \rangle \langle \ell m| .
\label{polar1}
\end{equation}

The coefficient, $\alpha_{\rm core}$, is the static dipole polarizability 
of the core and $g_{\ell}^2(r) = 1-\exp \bigl($-$r^6/\rho_{\ell}^6 \bigr)$
is a cutoff function designed to make the polarization potential finite
at the origin. The cutoff parameters, $\rho_{\ell}$, were tuned to
reproduce the binding energies of the Ga$^{2+}$ $ns$ ground state and
the $np$, $nd$ and $nf$ excited states. The Ga$^{3+}$ core polarizability
was chosen to be $\alpha_{\rm core} = 1.24$ $a_0^3$ \cite{johnson83a}.
The cutoff parameters for $\ell = 0 \to 3$ were 1.3074, 1.5235, 2.2035 
and 1.2977 $a_0$ respectively.

It is essential 
to include a two body polarization term, $V_{\rm p2}$, in the Hamiltonian  
to get accurate energy levels and polarizabilities for Ga$^+$. 
The polarization of the core by one electron is influenced by the 
presence of the second valence electron. Omission of the two-body term 
would typically result in a $4s^2$ $^1S^e_0$ state that would be too tightly 
bound. The importance of the two body polarization potential 
is discussed in ref.~\cite{norcross76a}. The two body polarization potential 
adopted for the present calculation has the form 
\begin{equation}
V_{\rm p2}({\bf r}_i,{\bf r}_j) = -\frac{\alpha_d} {r_i^3 r_j^3}
({\bf r}_i\cdot{\bf r}_j)g_{\rm p2}(r_i)g_{\rm p2}(r_j)\ ,
\label{polar2}
\end{equation}
where $g_{p2}$ has the same functional form as $g_{\ell}(r)$. The 
cutoff parameter for $g_{\rm p2}(r)$ was chosen as 1.583 $a_0$, the 
average of the cutoff parameters for $\ell = 0 \to 3$. Use of 1.583 
$a_0$ for the two-body cutoff parameter resulted in energies that 
were close to the experimental binding energies for most of the lowest 
lying states of Ga$^+$.  The current approach to solve the 
Schrodinger equation is termed as configuration interaction plus core 
polarization (CICP).   

There were a total of 195 valence orbitals with a maximum orbital
angular momentum of $\ell = 5$. The radial dependence of the orbitals 
were described by a mixture of STOs and Laguerre type orbitals
(LTOs) \cite{mitroy03f}. The number of active orbitals for 
$\ell = 0 \to 5$ were 50, 30, 30, 30, 30, and 25 respectively.
Some $\ell = 0$ valence orbitals were generated from the STOs 
used for the core. All the other orbitals were written as LTOs 
due to their superior linear dependence properties when compared 
with STO basis sets. The use of the large orbital basis resulted 
in wave functions and energies for the low-lying states that 
were close to convergence.  

The length of the CI expansions for the different states of Ga$^+$ 
ranged from 2000-7000. Some small changes were made to the $\rho_\ell$ 
values that were originally tuned to the Ga$^{2+}$ spectrum to improve 
the agreement of the Ga$^{+}$ energies with experiment. The oscillator 
strengths were computed with operators that included polarization 
corrections \cite{hameed72a,mitroy93a,mitroy03f}. The cutoff parameter 
used in the polarization correction to the dipole operator was 1.583 $a_0$.

\begin{table}[t]
\caption[]{ \label{energy}
Theoretical and experimental energy levels (in Hartree) for some of 
the low-lying states of the Ga$^{2+}$ and Ga$^{+}$ ions. The energies 
are given relative to the energy of the Ga$^{3+}$ core. The experimental 
energies for the spin-orbit doublets are averages with the usual 
$(2J+1)$ weighting factors.  The CICP energies for Ga$^+$ are those 
computed after additional tuning of the $\rho_{\ell}$ parameters.  
The experimental data were taken from the National Institute of Standards 
and Technology \cite{nistasd500}. }  
\begin{ruledtabular}
\begin{tabular}{lcc}
State & Experiment & CICP     \\ 
\hline
 \multicolumn{3}{c}{Ga$^{2+}$}      \\
$4s$ $^2S^e$ &  $-$1.1219500   & $-$1.1219503    \\
$4p$ $^2P^o$ &  $-$0.8269972   & $-$0.8269971    \\
$5s$ $^2S^e$ &  $-$0.4878602   & $-$0.4855755    \\
$4d$ $^2D^e$ &  $-$0.4723326   & $-$0.4723326    \\
$5p$ $^2P^o$ &  $-$0.3950115   & $-$0.3936100    \\
$4f$ $^2F^o$ &  $-$0.2842400   & $-$0.2842400    \\
$6s$ $^2S^e$ &  $-$0.2745341   & $-$0.2733390    \\
$5d$ $^2D^e$ &  $-$0.2668754   & $-$0.2663720    \\
$6p$ $^2P^o$ &  $-$0.2350878   & $-$0.2327966    \\
$5f$ $^2F^o$ &  $-$0.1818290   & $-$0.1819208    \\
$5g$ $^2G^e$ &  $-$0.1802746   & $-$0.1802675    \\
 \multicolumn{3}{c}{Ga$^{+}$}      \\
$4s^2$ $^1S^e_0$ &  $-$1.8830675   & $-$1.8830676       \\
$4s4p$ $^3P^o_0$ &  $-$1.6672451   & $-$1.6672449       \\
$4s4p$ $^1P^o_1$ &  $-$1.5609281   & $-$1.5609281       \\
$4s5s$ $^3S^e_1$ &  $-$1.4240174   & $-$1.4240171       \\
$4s5s$ $^1S^e_0$ &  $-$1.3970779   & $-$1.3999075       \\
$4p^2$ $^1D^e_2$ &  $-$1.3922558   & $-$1.3922558       \\
$4s4d$ $^3D^e_1$ &  $-$1.3644842   & $-$1.3644842       \\
$4p^2$ $^3P^e_1$ &  $-$1.3580662   & $-$1.3580660       \\
$4s5p$ $^3P^o_0$ &  $-$1.3434609   & $-$1.3433985       \\
$4s5p$ $^1P^o_1$ &  $-$1.3337993   & $-$1.3330456       \\
$4s4d$ $^1D^e_2$ &  $-$1.3081145   & $-$1.3113844       \\
$4s6s$ $^3S^e_1$ &  $-$1.2770280   & $-$1.2871275       \\
$4s6s$ $^1S^e_0$ &  $-$1.2736978   & $-$1.2781214       \\
$4p^2$ $^1S^e_0$ &  $-$1.2650309   & $-$1.2666402       \\
$4s5d$ $^3D^e_1$ &  $-$1.2581318   & $-$1.2581396       \\
$4s4f$ $^3F^o_2$ &  $-$1.2573356   & $-$1.2573356       \\
$4s4f$ $^1F^o_3$ &  $-$1.2572887   & $-$1.2571514       \\
\end{tabular}
\end{ruledtabular}
\end{table}

\section{Energies and Oscillator strengths}

\subsection{Energy levels} 

The energy levels of the present calculations are compared with 
experiment in Table \ref{energy}. The cut-off parameters
of the polarization potential were tuned to reproduce the 
experimental binding energies of the lowest states of each symmetry.
The energies of the lowest Ga$^{2+}$ states are all in agreement 
with experiment since this was the criteria used to tune the cutoff 
parameters. The excited states tend to under-bind the experimental 
energies by about 0.001-0.002 Hartree.

\begin{table}[tbh]
\caption[]{Absorption oscillator strengths for some low lying transitions of Ga$^{2+}$. The acronyms 
MP and RCC are defined in the text while RMBPT refers to relativistic many body perturbation theory 
and MCHF refers to (non-relativistic) multi-configuration Hartree-Fock.}
\label{ostrength1}
\begin{ruledtabular}
\begin{tabular}{lcccc}
  Transition  & CICP & MP \cite{victor83a}& RCC \cite{dutta13a} & Other \\ 
\hline  
$4s$$\to$$4p$ & 0.7959  & 0.915 & 0.8339 &  0.80000 (MP) \cite{curtis89a} \\  
              &         &       &        &  0.792 (MCHF) \cite{fischer77a} \\
	      &         &       &        &  0.8170  (RMBPT) \cite{chou97a}   \\
              &         &       &        & 0.843(5) Expt.~\cite{ansbacher85a} \\  
$4s$$\to$$5p$ & 0.00902 & 0.005 & 0.0066 &                                           \\
$4s$$\to$$6p$ & 0.00630 & 0.004 &        &                                            \\
$4p$$\to$$5s$ & 0.1511  & 0.140 & 0.1510 &                                            \\
$4p$$\to$$4d$ & 0.9317  & 1.000 & 0.9506 &  0.91348 (MP) \cite{curtis89a}          \\
$4p$$\to$$6s$ & 0.0220  & 0.020 & 0.0206 &                        \\
$4p$$\to$$5d$ & 0.0557  & 0.061 & 0.0592 &                       \\
$5s$$\to$$5p$ & 1.3745  & 1.354 & 1.3749 &                      \\
$5s$$\to$$6p$ & 0.00026 & 0     &        &                 \\
$4d$$\to$$5p$ & 0.2078  & 0.204 & 0.2092 &                       \\
$4d$$\to$$4f$ & 1.1163  & 1.113 & 1.1300 &                      \\
$4d$$\to$$6p$ & 0.00478 & 0.005 &        &                      \\
\end{tabular}
\end{ruledtabular}
\end{table}
Small adjustments to the cut-off parameters were made for the
calculations of the Ga$^+$ states. For example, the value of 
$\rho_0$ was reset to 1.2187 $a_0$ for the calculation of the 
states of the $^1S^e_0$ symmetry. The value of $\rho_0$ was 
fixed by requiring that the theoretical and experimental energies 
for the $4s^2 \ ^1S^e_0$ state be the same. Other fine tunings of the 
cut-off parameters were made for all symmetries. The most 
important levels for the calculation of the polarizabilities 
are the most tightly bound levels. The agreement between the 
theoretical and experimental energies for these levels minimises 
the impact that differences in the long range behaviour of the 
wave functions (which are influenced by the energy) 
will have on radial matrix elements that are part of the 
polarizability calculation.

\subsection{Oscillator strengths} 

The oscillator strengths for the transitions between the low lying 
states are listed in Table \ref{ostrength1} for Ga$^{2+}$ and Table
\ref{ostrength2} for Ga$^+$. The absorption oscillator strength from 
state $\psi_i$ to state $\psi_j$ is calculated according to the 
identity \cite{yan96a,mitroy03f},  
\begin{equation}
f^{(k)}_{ij} =  \frac {2 |\langle \psi_i;L_i \parallel \  r^k
{\bf C}^{k}({\bf \hat{r}}) \parallel \psi_{j};L_j \rangle|^2 \epsilon_{ji}}
{(2k+1)(2L_i+1)}  \ .
\label{fvaldef}
\end{equation}
In this expression, $\epsilon_{ji} = (E_j - E_i)$ is the energy 
difference between the initial state and final state, while $k$ is 
the polarity of the transition, and ${\bf C}^{k}({\bf \hat{r}})$
is a spherical tensor. Experimental energy differences were used 
for the calculation of oscillator strengths. The angular 
momentum weighted average energy difference were used for the 
Ga$^{2+}$ transitions. The energy differences of individual 
levels with the specific total angular momentum, $J$, were 
used for the triplet states of Ga$^+$.

\begingroup
\squeezetable 
\begin{table*}[tbh]
\caption[]{Absorption oscillator strengths for some low lying transitions of Ga$^{+}$. The 
RRPA column lists calculations performed in the relativistic random phase approximation.  
The CICP oscillator strengths were obtained from calculations where the core-polarization 
potentials were tuned to give the energies of the identified spin-orbit state.   
}
\label{ostrength2}
\begin{ruledtabular}
\begin{tabular}{lccccccc}
Transition  & CICP  & MP \cite{victor83a} &  MCDF \cite{jonsson06a} & CI  & RRPA & MCHF &  Experiment\\ 
\hline  
$4s^2 \ ^1S^e_0$$\to$$4s4p$ $^1P^o_1$ &1.7227 & 1.890 &1.71 &1.76 \cite{tayal91a}  &1.8620 \cite{chou94a} & 1.75 \cite{fischer78a}& 1.85(15) \cite{andersen79a}\\
& &&&1.827 \cite{hibbert95a} &1.983 \cite{huang92a} & 1.71 \cite{fischer79a}&      \\
& && & 1.704 \cite{mcelroy05a}   &1.691 \cite{shorer78a} &                       &                \\
$4s^2 \ ^1S^e_0$$\to$$4s5p$ $^1P^o_1$ & 0.00478& 0.017 & 0.00546& 0.0066 \cite{mcelroy05a} &0.00683 \cite{shorer78a}& & \\
$4s^2$ $^1S^e_0$$\to$$4s4p$ $^3P^o_1$ &   &  &5.92[-4] &8.1[-4] \cite{hibbert95a} & 4.7534[-4] \cite{chou94a}&& \\ 
 & &&&6.37[-4] \cite{mcelroy05a} & 3.803[-4] \cite{huang92a}&         &                 \\
$4s4p \ ^1P^o_1$$\to$$4s5s$ $^1S^e_0$ & 0.1403 & 0.154 & 0.149  & 0.141 \cite{mcelroy05a}  &  &   &     \\
$4s4p \ ^1P^o_1$$\to$$4s6s$ $^1S^e_0$ &0.00275 & 0.003 & &       &      &         &         \\
$4s4p \ ^1P^o_1$$\to$$4p^2$ $^1S^e_0$ &0.2051  & 0.222 & &       &      &         &          \\
$4s4p$ $^1P^o_1$$\to$$4p^2$ $^1D^e_2$ &0.0120 & 0.005 &0.0172 & 0.029 \cite{tayal91a}&&0.58[-3] \cite{fischer79a}&   \\
$4s4p \ ^1P^o_1$$\to$$4s4d$ $^1D^e_2$ &1.2201 &       &1.31  & 1.08 \cite{tayal91a} & & 1.21 \cite{fischer79a} &    \\
$4s4p \ ^1P^o_1$$\to$$4s5d$ $^1D^e_2$ &0.3441 &0.417 &  &         &  &                    \\
\hline 
$4s4p \ ^3P^o_0$$\to$$4s5s$ $^3S^e_1$ &0.1335 &0.136 & 0.142 &      &        &         &        \\ 
$4s4p \ ^3P^o_0$$\to$$4s6s$ $^3S^e_1$ &0.0195 &0.020 &0.0179 &   &        &         &        \\ 
$4s4p \ ^3P^o_0$$\to$$4s4d$ $^3D^e_1$ &0.7946 &0.830 &0.81 &   &        &         &        \\ 
$4s4p \ ^3P^o_0$$\to$$4s5d$ $^3D^e_1$ &0.1039 &0.112 &0.0958 &     &        &         &        \\ 
$4s4p \ ^3P^o_0$$\to$$4p^2$ $^3P^e_1$ &0.5683 &0.630 &0.569 &  &   &        &         \\ 
$4s5s \ ^3S^e_1$$\to$$4s5p$ $^3P^o_0$ &1.5090 &1.319  &1.303 &      &        &             &    \\ 
\end{tabular}
\end{ruledtabular}
\end{table*}
\endgroup
There have been a number of calculations of the energy 
levels and oscillator strengths for Ga$^{2+}$ 
\cite{victor83a,owono05a,owono97a,curtis89a,chichkov81a,fischer77a,ansbacher85a}.   
Not all of the calculations of oscillator strengths have been tabulated. 
Table \ref{ostrength1} gives oscillator strengths that are deemed to 
be the most accurate or of particular relevance to the present calculations.
The most comprehensive calculations for Ga$^{2+}$ appear to be a model 
potential (MP) calculation \cite{victor83a} and  most recently 
a relativistic coupled cluster (RCC) calculation \cite{dutta13a}. 
The reliability of the MP calculation from \cite{victor83a} is 
questionable since the oscillator strength for the resonance $4s \to 4p$ 
transition is at variance with the CICP calculation and experiment. 
The CICP and RCC give oscillator strengths that mostly lie within 5$\%$ 
of each other with the exceptions occurring for oscillator strengths 
that are small.  

The oscillator strengths reported in Table \ref{ostrength2} for Ga$^+$ were 
taken from a variety of sources 
\cite{andersen79a,hibbert95a,huang92a,fischer79a,mcelroy05a,shorer78a,victor83a,jonsson06a}.  
There is a good deal of variety in values of the oscillator strengths for the resonance 
$4s^2$ $^1S^e_0$$\to$$4s4p$ $^1P^o_1$ transition, with values ranging from 
1.69 to 1.983. However, two of the most recent calculations, a multi-configuration 
Dirac-Fock (MCDF) and a configuration interaction (CI) calculation gave values of 
1.71 \cite{jonsson06a} and 1.704 \cite{mcelroy05a} respectively.  These are at a  
1$\%$ level of agreement with the CICP oscillator strength.  The MCDF calculation explicitly allowed for core-valence correlation, but only allowed for 
excitations from the $3d^{10}$ orbital.  The more tightly bound orbitals 
account for about 20$\%$ of the core polarizability.    
The oscillator 
strength for the $4s^2$ $^1S^e_0$$\to$$4s4p$ $^3P^o_1$ inter-combination transition 
is small and will not make a significant contribution to the polarizability.   

\begin{table}[t]
\caption[]{ \label{fcore}
The pseudo-oscillator strength distribution for the Ga$^{3+}$ ion core. The
dipole polarizability of the core is 1.24 a.u. \cite{johnson83a} while the 
quadrupole polarizability was 2.345 a.u. \cite{johnson83a}.}  
\begin{ruledtabular}
\begin{tabular}{lcc}
\multicolumn{1}{c}{} & \multicolumn{1}{c}{$\epsilon_i$}& \multicolumn{1}{c}{$f^{(k)}_i$} \\
\hline   
dipole     & 380.70060 & 2.0     \\
           &  50.04884 & 2.0     \\
           &  44.37403 & 6.0     \\
           &   8.27546 & 2.0     \\
           &   6.36168 & 6.0     \\
           &   3.07331 & 10.0    \\
quadrupole & 380.35370 & 0.01308  \\
           &  49.70194 & 0.22940 \\
           &  44.02713 & 0.54180 \\
           &  7.92856 & 2.00991 \\
           &   6.01478 & 6.61121  \\
           &   2.72641 &15.83296 \\
\end{tabular}
\end{ruledtabular}
\end{table}

Transitions originating on the $4s4p$ $^3P^o_0$ multiplet will determine the 
polarizability of this state. The overall level of agreement between the present 
CICP oscillator strengths and those of the MCDF calculation \cite{jonsson06a} 
is good, with only a 2$\%$ difference in oscillator strengths for the two 
strongest transitions.  There is a 6$\%$ disagreement for the 
$4s4p$ $^3P^o_0$ $\to$ $4s5s$ $^3S^e_1$ transition, but this oscillator 
strength is small and it only makes a $10\%$ contribution to the  
polarizability of the $4s4p$ $^3P^o_0$ state.  There is no explicit statement 
regarding the size of the orbital space used in the MCDF calculation, 
but it is likely to be significantly smaller than that used for the  
present calculations.   

\section{Polarizabilities and BBR shifts}

\subsection{Scalar and tensor polarizabilities} 

This analysis is done under the assumption that spin-orbit effects 
are small and the radial parts of the wave functions are the 
same for the states with different $J$. All the polarization 
parameters reported here are calculated using  
oscillator strength sum rules. The multipole oscillator strengths 
$f^{(k)}_{ij}$ are defined in Eq.~(\ref{fvaldef}). Then the 
adiabatic multipole polarizabilities $\alpha_{k}$ from 
the state $i$ are written as \cite{mitroy03e,mitroy10a}
\begin{equation}
\alpha_{k} = \sum_j \frac{f^{(k)}_{ij}}{\epsilon^2_{ji}} \ .
\label{alphav}
\end{equation}
A related sum rule is the non-adiabatic multipole polarizability 
$\beta_{k}$ \cite{dalgarno68a,mitroy03f}, which is defined as 
\begin{equation}
\beta_{k} = \frac{1}{2}\sum_j \frac{f^{(k)}_{ij}}{\epsilon^3_{ji}} \ . 
\label{betav}
\end{equation}
This is useful for the analysis of resonant excitation stark ionization 
spectroscopy (RESIS) \cite{lundeen05a} experiments. A RESIS experiment 
would be able to determine the polarizabilities of Ga$^+$ and Ga$^{2+}$ 
to better than 1$\%$ accuracy. 

The dynamic polarizability to lowest order variations in the 
frequency can be written \cite{mitroy10a} as    
\begin{equation}
\alpha_{k}(\omega) \approx \alpha_{k}(0) + \omega^2 S_{k}(-4) + \ldots,  
\label{sum2}
\end{equation}
where $S_k(-4)$ is  
\begin{equation}
S_{k}(-4) = \sum_j \frac{f^{(k)}_{ij}}{\epsilon^4_{ji}} \ .
\label{sum1}
\end{equation}

States with a non-zero angular momentum will also have a tensor 
polarizability \cite{angel68a,zhang07a,mitroy10a}. For a state with angular 
momentum $L_0$($J_0$), this is defined as the polarizability of the 
magnetic sub-level with $M = L_0$($M = J_0$). The total polarizability 
is written in terms of both a scalar and tensor polarizability.
The scalar polarizability represents the average shift of the
different $M$ levels while the tensor polarizability gives the
differential shift.

\begin{table*}[t]
\caption[]{ \label{polar}
The polarizabilities of some low lying states of the Ga$^{2+}$ and 
Ga$^{+}$ ions. 
The scalar adiabatic polarizabilities, $\alpha_{k}$, are listed 
along with some non-adiabatic, $\beta_{k}$, and tensor,  
$\alpha^{(1)}_{2, L_0 L_0}$, polarizabilities.  
The polarizabilities are in atomic units and the notation 
$a[b]$ means $a\times10^b$. }  
\begin{ruledtabular}
\begin{tabular}{lccccccc}
\multicolumn{1}{c}{State} &  
\multicolumn{1}{c}{$\alpha_1$ }&$S_1(-4)$ & \multicolumn{1}{c}{$\beta_1$ }& \multicolumn{1}{c}{$\alpha^{(1)}_{2, L_0 L_0}$ } & \multicolumn{1}{c}{$\alpha_2$ }& \multicolumn{1}{c}{$\beta_2$ }&  \multicolumn{1}{c}{$\alpha_3$} \\
\hline
          \multicolumn{8}{c}{Ga$^{2+}$}   \\
4s $^2S^e$ &  10.027   & 95.555    & 14.643   &  0        &  24.153   & 16.764 & 141.35     \\
4p $^2P^o$ &  7.4274   & 39.282    & 17.657   &  0.69722  &  46.325   & 40.977 & 752.61      \\
5s $^2S^e$ &  156.85   & 1.8461[4] & 864.69   &  0        &  9.9017[3]& 2.9156[4]& 5.0411[4]     \\
4d $^2D^e$ &  64.830   & 6.6898[3] & 317.93   & $-$39.912 & $-$1.4309[3] & 5.8572[4]& 1.8759[4]  \\
5p $^2P^o$ & $-$12.784 & $-$9.9801[3] & 1.0365[3] &  31.986 &  3.4326[3] & 1.3990[4]& 1.3351[5]  \\
           \multicolumn{8}{c}{Ga$^{+}$}   \\
$4s^2$ $^1S^e_0$ & 17.946 & 160.10 & 26.005 &   0       & 80.661 & 73.978 & 623.96    \\
$4s4p$ $^3P^o_0$ & 19.576 & 201.96 & 30.330 &   0   & 116.02 & 141.92 & 879.34  \\
$4s4p$ $^1P^o_1$ & 28.858 & 533.19 & 76.423 & $-$4.8720 & 226.25 & 348.70 & 2.2791[3]   \\           
\end{tabular}
\end{ruledtabular}
\end{table*}

This tensor polarizability can be expressed in terms of $f$-value sum 
rules. For an $L_0 = 1$ initial state, one can write the tensor 
polarizability for a dipole field as \cite{zhang07a,mitroy10a}
\begin{equation}
\alpha_{2,L_0L_0} = -\biggl( \ \sum_{n,L_n=0} \frac {f_{0n} } 
{\epsilon_{n0}^2} -\frac{1}{2} \sum_{n,L_n=1} \frac {f_{0n} } 
{\epsilon_{n0}^2} + \frac{1}{10} \sum_{n,L_n=2} \frac {f_{0n} } 
{\epsilon_{n0}^2} \biggr) \ .
\label{alpha2L}
\end{equation}
The core does not make a contribution to the tensor polarizability.  
Expressions for the general tensor polarizabilities have been given 
elsewhere  \cite{zhang07a}.

\subsubsection{The polarizability of the Ga$^{3+}$ core} 

The energy distribution of the oscillator strengths originating from 
core excitations was estimated using a semi-empirical technique 
\cite{mitroy03f}. This approach utilizes $f$-value sum rules and 
identities to construct the pseudo-oscillator strength distributions. 
The sum rules and identities are    
\begin{eqnarray}
f^{k}_i &=& k N_i \langle r^{2k-2}_{i} \rangle , \nonumber \\   
\alpha_{{\rm k},{\rm core}} &=& \sum_{i \in {\rm core}} \frac {f^{k}_i} {(\epsilon_i)^2} \ .  
\label{alphacore}
\end{eqnarray}
In these expressions, $N_i$ is the number of electrons in a core orbital, 
and $\langle r^{2k-2}_{i} \rangle$ is a radial expectation value of the orbital.  
The $\epsilon_i$ is initially set to be the single-particle (Koopmans) energy of 
the HF orbitals. They are then shifted by an additive constant, e.g.  
$\epsilon_i = \epsilon_{\rm HF} + \Delta$, and the parameter $\Delta$ is adjusted 
until the computed core polarizability is equal to an estimate of the core polarizability  
obtained from another source \cite{johnson83a}. The pseudo-oscillator strength 
distribution used for the Ga$^{3+}$ core is given in Table \ref{fcore}. This 
distribution was used in the determination of all oscillator strength sum rules. 

\subsubsection{Polarizabilities} 

Table \ref{polar} gives the multipole polarizabilities 
of the lowest five states of the Ga$^{2+}$ ion and the lowest three 
states of the Ga$^{+}$ ion.  The energies of the lowest lying states in 
the Ga$^{2+}$ polarizability calculations were adjusted to be the same as 
the spin-orbit averaged experimental energies listed in Table \ref{energy}. 
The polarizabilities of excited states are more sensitive to small 
errors in calculated energies since the energy differences can be 
much smaller.  

Energy adjustments were made when performing the polarizability calculations 
of the Ga$^+$ ion states.  First, for the singlet states, the energies 
of the lowest excited states were adjusted to be the same as the experimental
binding energies.  The purpose of the triplet state calculations was to determine 
the polarizability of the $4s4p \ ^3P^o_0$ state.  The cutoff parameters for the 
core-polarization potential were adjusted so that the CICP $4s4p \ ^3P^o$ state 
energy was the same as the experimental $4s4p \ ^3P^o_0$ state energy.  Further, 
the cutoff parameters for other symmetries were adjusted so that the 
excited state energies were those of the spin-orbit states that could undergo 
a direct multipole transition with the $4s4p \ ^3P^o_0$ state.  For example, 
the parameters were tuned so that the $4p^2 \ ^3P^e$ and $4s4d \ ^3D^e$ excited 
state energies were set to be those of the $J = 1$ state, and the energies of 
the $^3F^o$ states were set to be those of the $J = 2$ state.  In effect, 
the CICP matrix elements were calculated using wave functions that have the 
energies of the appropriate spin-orbit states.  Exact agreement between the 
tuned CICP energies and the experimental energies was only achieved for the 
lowest energy state of each symmetry. For the second and third excited state 
of each symmetry, CICP matrix elements without any further adjustment were 
used with the experimental binding energies of the appropriate spin-orbit component.

The tensor polarizabilities and non-adiabatic polarizabilities 
as well as the related sum rules $S_{k}(-4)$ of these states are also listed in
Table \ref{polar}.  The contributions from different transitions
to the dipole polarizabilities of the $4s^2$ $^1S^e_0$, $4s4p$ $^3P^o_0$, and
$4s4p$ $^1P^o_1$ states are detailed in Table \ref{breakdown}.  The 
$4s4p$ $^3P^o_0$ does not have a tensor polarizability since it is 
the $J = 0$ spin-orbit component.   

The $4s^2$ $^1S^e_0$ ground state polarizability is dominated by the resonant 
transition which contributes about 92$\%$ of the polarizability (refer to 
Table \ref{breakdown}). The next most significant 
contribution to the polarizability comes from the Ga$^{3+}$ core. The 
uncertainty in the CICP line strength for the resonant transition is 
assessed to be $\pm 2\%$. This was based on the variation between the 
CICP, MCDF \cite{jonsson06a} and CI \cite{mcelroy05a,tayal91a} oscillator strengths 
for this transition. The uncertainty in the RRPA core polarizability 
of 1.24 is assessed to be $\pm 1\%$. This uncertainty is based on an 
estimate of the uncertainty in the core polarizability of Ca$^{+}$ 
\cite{safronova11a}. The total uncertainty in the ground state 
dipole polarizability of 17.95 is 0.34 a.u..      

There has been an estimate of the Ga$^{+}$ dipole polarizability by 
using oscillator strength sum rules and regularities in the 
$4s^2 \ ^1S^e_0 \to 4s4p \ ^1P^o_1$ line strengths between members of 
the isoelectronic series \cite{reshetnikov08a}. They report a value 
of 18.14(44) a.u.. This polarizability appears to be for the valence 
only part of the polarizability.

\begin{table}[t]
\caption[]{ \label{breakdown}
Breakdown of the contributions to the dipole polarizabilities of the Ga$^{+}$
clock transition states. The $\delta \alpha_1$ column gives the contribution
from the indicated transition class. The $\sum \alpha_1$ column gives
the accumulated sum. The final polarizabilities are given in bold-face and
the uncertainties in the last digits are given in brackets.
}
\begin{ruledtabular}
\begin{tabular}{lcc}
\multicolumn{1}{c}{Transition(s)}& \multicolumn{1}{c}{$\delta \alpha_1$} & \multicolumn{1}{c}{$\sum$ $\alpha_1$} \\
\hline   
\multicolumn{3}{c}{$4s^2$ $^1S^e_0$ state}    \\
$4s^2$ $^1S^e_0$ $\to$ $4s4p$ $^1P^o_1$  & 16.6010 & 16.6010     \\
$4s^2$ $^1S^e_0$ $\to$ $4s5p$ $^1P^o_1$  & 0.0158  & 16.6168     \\
$4s^2$ $^1S^e_0$ $\to$ $nP$ $^1P^o_1$    & 0.0891  & 16.7059    \\
Core                                     &  1.24   &  {\bf 17.95(34)}  \\
\multicolumn{3}{c}{$4s4p$ $^3P^o_0$ state}    \\
$4s4p$ $^3P^o_0$ $\to$ $4s5s$ $^3S^e_1$  & 2.2565  &  2.2565  \\
$4s4p$ $^3P^o_0$ $\to$ $nS$ $^3S^e_1$    & 0.2894  &  2.5459  \\
$4s4p$ $^3P^o_0$ $\to$ $4p^2$ $^3P^e_1$  & 5.9447  &  8.4900  \\
$4s4p$ $^3P^o_0$ $\to$ $nP$ $^3P^e_1$    & 0.0124  &  8.5030  \\
$4s4p$ $^3P^o_0$ $\to$ $4s4d$ $^3D^e_1$  & 8.6681  & 17.1711  \\
$4s4p$ $^3P^o_0$ $\to$ $nD$ $^3D^e_1$    & 1.1644  & 18.3355  \\
Core                            &  1.24   & {\bf 19.58(38)}    \\
\multicolumn{3}{c}{$4s4p$ $^1P^o_1$ state}    \\
$4s4p$ $^1P^o_1$ $\to$ $4s^2$ $^1S^e_0$  &$-$5.5337&$-$5.5337  \\
$4s4p$ $^1P^o_1$ $\to$ $4s5s$ $^1S^e_0$  & 5.2257  &$-$0.3080  \\
$4s4p$ $^1P^o_1$ $\to$ $4p^2$ $^1S^e_0$  & 2.3426  &  2.0346  \\
$4s4p$ $^1P^o_1$ $\to$ $nS$ $^1S^e_0$    & 0.3269  &  2.3615  \\
$4s4p$ $^1P^o_1$ $\to$ $4p^2$ $^1D^e_2$  & 0.4203  &  2.7818  \\
$4s4p$ $^1P^o_1$ $\to$ $4s4d$ $^1D^e_2$  & 19.0891 & 21.8709  \\
$4s4p$ $^1P^o_1$ $\to$ $4s5d$ $^1D^e_2$  & 3.4815  & 25.3524  \\
$4s4p$ $^1P^o_1$ $\to$ $nD$ $^1D^e_2$    & 2.2400  & 27.5924  \\
$4s4p$ $^1P^o_1$ $\to$ $4p5p$ $^1P^e_1$  & 0.0002  & 27.5926  \\
$4s4p$ $^1P^o_1$ $\to$ $nP$ $^1P^e_1$    & 0.0249  & 27.6175    \\
Core                                     & 1.24& {\bf 28.86(3.36)}    \\
\end{tabular}
\end{ruledtabular}
\end{table}

The CICP calculation of the ground state polarizability did not take 
into consideration the contribution from the $4s^2 \ ^1S^e_0$ $\to$ 
$4s4p \ ^3P^o_1$ transition. The oscillator strength for this 
transition is only $6.0 \times 10^{-4}$ \cite{jonsson06a}, so this
transition can be safely omitted from the determination of the 
polarizability. This also justifies the omission of the spin-orbit 
interaction from the effective Hamiltonian for the valence electrons.  

The $4s4p \ ^3P^o_0$ state polarizability was computed to be 19.58 
a.u.. Table \ref{breakdown} details the contributions of different 
transitions to this polarizability. The excitations to the lowest 
three states make a contribution of 86$\%$ to the total polarizability 
with the remainder being split in a roughly equal manner between the 
core and higher valence excitations. The error analysis for this 
polarizability assumed a 6$\%$ uncertainty in the 
$4s4p \ ^3P^o_0 \to 4s5s \ ^3S^e_1$ oscillator strength, and 1$\%$ 
uncertainties in the $4s4p \ ^3P^o_0 \to 4p^2 \ ^3P^e_1$ and 2$\%$ uncertainties in the
$4s4p \ ^3P^o_0 \to 4s4d \ ^3D^e_1$ oscillator strengths. Adding in the 
uncertainty for the core polarizability gives a net uncertainty of 0.38 a.u.. 

The present CICP $4s4p \ ^1P^o_1$ polarizability is 28.86 a.u. and the 
most significant contribution comes from $4s4p \ ^1P^o_1 \to 4s4d \ ^1D^e_2$
transition which contributes 66$\%$ of the total polarizability. The 
contributions from $4s4p \ ^1P^o_1 \to 4s^2 \ ^1S^e_0$ and
$4s4p \ ^1P^o_1 \to 4s5s \ ^1S^e_0$ transitions almost cancel each other.
The $4s4p \ ^1P^o_1$ state has a larger relative uncertainty than the 
other two states. First, this state has both positive and negative 
contributions to the polarizability. Second, there are a number of 
strongly interacting configurations, e.g the $4p^2 \ ^1D^e$ 
and $4snd \ ^1D^e$ configurations, which make the calculations of 
the matrix elements for these states more sensitive to the final
details of the structure model.

Based on the variation between the CICP and MCDF \cite{jonsson06a} oscillator 
strengths of the most important transitions, we set the uncertainties as follows: 
1$\%$ for the $4s4p \ ^1P^o_1 \to 4s^2 \ ^1S^e_0$ transition, 
6$\%$ for the $4s4p \ ^1P^o_1 \to 4s5s \ ^1S^e_0$ transition, 
8$\%$ for the $4s4p \ ^1P^o_1 \to 4p^2 \ ^1S^e_0$, $4s4p \ ^1P^o_1 \to nS \ ^1S^e_0$, 
and $4s4p \ ^1P^o_1 \to 4s4d \ ^1D^e_2$ transitions, and 20$\%$ for other transitions. 
Consideration of all the uncertainties in these transitions, and the uncertainty 
in the core polarizability gives a total uncertainty of 3.36 a.u. for the 
$4s4p \ ^1P^o_1$ state.

\subsection{The BBR shift}

The BBR shift (in Hz) can be written as
\begin{equation}
\Delta \nu_{\rm BBR} = 6.579684 \times 10^{15} \left(
\Delta E_{\rm upper} - \Delta E_{\rm lower} \right),
\label{BBR3}
\end{equation}
where the electric dipole (E1) induced BBR energy shift of an
atomic state can be approximately calculated as \cite{porsev06a,mitroy10a}
\begin{equation}
\Delta E \approx -\frac{2}{15} (\alpha \pi )^3 \alpha_1(0) T^4  \ .
\label{BBR1}
\end{equation}
The dipole polarizability of the relevant quantum state is $\alpha_1$
and $T$ is the temperature. In this expression the temperature in K 
is multiplied by 3.1668153 $\times$ 10$^{-6}$. Knowledge of the dipole
polarizabilities permits a temperature dependent BBR correction to be 
made to the clock. The uncertainty in the E1 BBR shift can be written
as
\begin{equation}
\delta (\Delta \nu_{\rm BBR}) = \Delta \nu_{\rm BBR} \left( \frac{\delta (\Delta \alpha_1)} {\Delta \alpha_1}
+ \frac{4 \delta T} {T} \right).
\label{BBR4}
\end{equation}
Using the CICP polarizabilities and setting  
$T = 300$ K gives $\nu_{4s^2 \ ^1S^e} =$ $-$$0.1545\pm0.0029$ Hz and 
$\nu_{4s4p \ ^3P^o_0} =$ $-$$0.1686\pm0.0033$ Hz. 

In the CICP calculation the dipole polarizability difference 
for the $4s^2$ $^1S^e_0$ $\to$ $4s4p$ $^3P^o_0$ clock transition is 
$\Delta \alpha_1= 1.63\pm 0.72 \ a_0^3$. Using this value of 
$\Delta \alpha_1$ leads to a net frequency shift at 300 K of 
$\Delta \nu= -0.0140\pm0.0062$ Hz. 

A small correction to the polarizabilities needs to be considered to 
potentially allow for a slight variation due 
to the finite temperature of the BBR field,    
\begin{equation}
\alpha_1(T)=\alpha_1(1+\eta) \ .  
\end{equation}
The factors, $\alpha_1(T)$, is the polarizability after correction, and 
$\eta$ is the dynamic correction factor. The leading order term of $\eta$ 
is given by \cite{porsev06a,mitroy09b}
\begin{equation}
\eta \approx  -\frac{40 \pi^2 T^2}{21 \alpha_1(0)} S_1(-4) \ .
\label{eta}
\end{equation}
The value of $\eta$ was found to be quite small. In the present CICP 
calculation, it was $-$$1.51 \times 10^{-4}$ for the $4s^2$ $^1S^e_0$
state and $-$$1.85 \times 10^{-4}$ for the $4s4p$ $^3P^o_0$ state.
The net change in the frequency due to these two corrections would be 
$(2.3 \times 10^{-5} - 3.1 \times 10^{-5}) = -8 \times 10^{-6}$ Hz. 
This change in frequency is much smaller than the uncertainty in the 
BBR frequency shift.    

There is one transition rate that is relevant to the operation 
of a Ga$^+$ optical frequency standard, namely the rate of the 
$4s^2 \ ^1S^e \to 4s4p \ ^3P^o_0$ transition.  A MCDF calculation 
has obtained the value of 0.334 s$^{-1}$ \cite{jonsson06a}.  The 
natural line-width of the clock transition is 0.053 Hz.     

\begin{table}[t]
\caption[]{ \label{summary}
Parameters of the $ns^2 \ ^1S^e_0 \to nsnp \ ^3P^o_0$ clock transitions 
for the group III ions.  The BBR shifts are evaluated at 300 K.   
The $\nu_{\rm nat}$ column gives available data for the 
natural line-width of the clock transitions.}
\begin{ruledtabular}
\begin{tabular}{lcccc}
Ion &  Method  & $\Delta \alpha$  (a.u.) & $\nu_{\rm BBR}$ (Hz) & $\nu_{\rm nat}$ (Hz)  \\
\hline   
B$^+$ &  CICP  \cite{cheng12a} & $-$1.86(6) & 0.0160(5)   &     \\
      &  CI+MBPT   \cite{safronova11a} & $-$1.85(18)  & 0.0159(16) &  \\
Al$^+$ &  CICP     \cite{mitroy09b}  & 0.48(37) & $-$0.0042(32) &       \\
       &  CI+MBPT  \cite{safronova11c}  & 0.495(50)    &$-$0.00426(43)  &        \\
       &  Exp. \cite{rosenband07a} &  &  & 0.0077 \\ 
Ga$^+$ &  CICP                        &   1.63(72) &  $-$0.0140(62)    &    \\
      &  MCDF \cite{jonsson06a} &          &             & 0.053 \\
In$^+$ &  CI+MBPT     \cite{safronova11c}     & 2.01(20) &$-$0.0173(17) &  \\
       &  Exp. \cite{becker01a} &  & & 0.82   \\ 
Tl$^+$ &  CI+MBPT   \cite{zuhrianda12a}   &  1.83(18)  &  $-$0.0157(16) & \\
\end{tabular}
\end{ruledtabular}
\end{table}

\section{Conclusions}

The dipole and quadrupole polarizabilities of the $4s$, $4p$, 
$4d$, $5s$ and $5p$ states of Ga$^{2+}$ have been determined by 
diagonalizing the effective Hamiltonian in a large basis. The 
dipole and quadrupole polarizabilities of the $4s^2 \ ^1S^e_0$,
$4s4p \ ^3P^o_0$, and $4s4p \ ^1P^o_1$ states of Ga$^+$ have 
been determined from large dimension CI calculations.  

The BBR shift for the $4s^2 \ ^1S^e_0 \to 4s4p \ ^3P^o_0$ clock transition 
has been determined to be $-$0.0140(62) Hz. The negative value means 
the frequency of the clock transition is reduced by BBR effects. 
The dynamic correction to the BBR shift has been found to 
be negligible at the level of precision used in this manuscript. The 
very small BBR shift is consistent with the small values reported 
for other group III ions
\cite{rosenband06b,mitroy09b,safronova11c,zuhrianda12a,cheng12a}.  
Table \ref{summary} is a summary table of polarizability differences and BBR 
shifts for the clock transitions of the group III ions.  The main 
trend is for the natural linewidth to steadily increase for the 
heavier atoms while the BBR shift stays remains relatively small.

\begin{acknowledgments}

The work was supported by the Australian Research Council Discovery 
Project DP-1092620. Dr Yongjun Cheng was supported by a grant from 
the Chinese Scholarship Council.

\end{acknowledgments}

\end{document}